\documentclass[
amsmath,amssymb,
aps,
floatfix,
]{revtex4-2}

\usepackage{graphicx}

\usepackage{dcolumn}
\usepackage{bm}
\usepackage{hyperref}

\usepackage{color}
\usepackage{subcaption}

\newcommand{\GM}{\textcolor{blue}}

\begin{document}

\def\be{\begin{equation}}
\def\ee{\end{equation}}

\def\d{\mbox{\rm d}}
\def\LCDM{$\Lambda$CDM\,}

\title{The Bondi universe: Can negative mass drive the cosmological expansion?}

\author{Giovanni Manfredi}
\email{giovanni.manfredi@ipcms.unistra.fr}
\affiliation{Universit\'e de Strasbourg, CNRS, Institut de Physique et Chimie des Mat\'eriaux de Strasbourg, UMR 7504, F-67000 Strasbourg, France}
\author{Jean-Louis Rouet}
\affiliation{Universit\'e d'Orl\'eans, CNRS/INSU, BRGM, ISTO, UMR7327, F-45071 Orl\'eans, France}
\author{Bruce Miller}
\affiliation{Department of Physics and Astronomy, Texas Christian University, Fort Worth, TX 76129}

\date{\today}

\begin{abstract}

We identify a new cosmological coincidence that parallels the well‑known matter/dark‑energy coincidence: the present‑epoch transition of the universe from a weakly coupled (collisionless) to a strongly coupled (collisional) gravitational regime. Within a cosmological model containing equal amounts of positive and negative Bondi masses -- consistent with the weak equivalence principle and momentum conservation -- we show that this coupling transition naturally coincides with the shift from a coasting to an accelerating expansion. A linear response analysis of the corresponding Vlasov-Poisson system reveals that mixed positive-negative mass configurations are always unstable, with growth rates that increase at shorter wavelengths, thereby driving the system toward strong coupling. Using long‑time, exact one-dimensional N-body simulations, we demonstrate that the universe undergoes three successive expansion phases: an initial ballistic regime, an intermediate random‑walk acceleration driven by sporadic Bondi encounters, and finally a uniformly accelerating phase triggered by the formation of stable positive/negative mass pairs. The onset of this last phase occurs precisely when the coupling parameter crosses unity, linking the two cosmological coincidences through a single dynamical mechanism. These results suggest that cosmic acceleration may arise from the nonlinear dynamics of a gravitationally neutral mixed‑mass universe, without invoking dark energy or a cosmological constant.

\end{abstract}

\maketitle

\section{Introduction}\label{sec:intro}

Over the past two decades, the Standard Cosmological Model ($\Lambda \rm CDM$) has evolved into a robust framework that accurately accounts for a wide spectrum of cosmological phenomena \cite{Frieman2008}. It provides consistent explanations for primordial nucleosynthesis, the cosmic microwave background (CMB) radiation, baryonic acoustic oscillations (BAOs), and the luminosity distances of type-1a supernovae (SN1a). Although the reliability of individual datasets varies, their collective interpretation strongly indicates that the universe contains far more than just conventional (baryonic) matter. In recent years, however, notable discrepancies have arisen between measurements of the Hubble constant $H_0$ derived from early-universe observations and those obtained from the local universe \cite{Riess2020}. These differences, reaching a statistical significance of nearly $5\sigma$, are difficult to attribute solely to systematic uncertainties, though it remains uncertain whether they herald the emergence of new physics. 
Moreover, the latest results from the Dark Energy Survey appear to indicate a potential decline in the effective dark‑energy contribution, which, if confirmed, would point toward physics that cannot be captured by a pure cosmological constant \cite{DES2025,DES-PRD,Camilleri2024}.

Despite its impressive descriptive power, the $\Lambda \rm CDM$ model is not a complete theory. According to its predictions, baryonic matter -- the familiar particles governed by the Standard Model of particle physics and routinely observed in laboratory experiments -- makes up less than 5\% of the universe’s total mass-energy budget. The remainder consists of cold dark matter (CDM, approximately 25\%) and dark energy, represented by a cosmological constant $\Lambda$ (roughly 70\%). 
This situation has spurred extensive research efforts, both experimental (aimed at detecting dark matter) and theoretical (exploring alternative scenarios that might eliminate the need for dark matter, dark energy, or both).

One of the conceptual challenges within the framework of the $\Lambda \rm CDM$ model is the so-called ``coincidence problem", which arises from the observation that the energy densities of matter and dark energy are of the same order of magnitude in the current epoch, despite their vastly different evolutionary histories. In standard cosmology, the energy density of matter scales as $a^{-3}$, where $a(t)$ is the scale factor of the expanding universe, while the energy density associated with the cosmological constant $\Lambda$ remains constant over time. Given these distinct scaling behaviors, the fact that their contributions to the total energy budget coincide precisely at the present epoch appears as an unlikely coincidence. This apparent fine-tuning raises questions about the underlying physical mechanisms that could account for such a temporal alignment, and has motivated various theoretical proposals, including dynamical dark energy models and anthropic reasoning within the context of a multiverse \cite{Velten2014,Oscoz2008,Shimon2024}. However, none of these approaches has yet provided a universally accepted resolution to the problem.

Some authors have questioned the empirical basis for the claimed acceleration of the universe's expansion, suggesting instead that observational data may be consistent with a ``coasting" cosmological model, in which the expansion rate remains constant over time, exhibiting neither acceleration nor deceleration. This scenario closely resembles the model originally proposed by E.~A.~Milne \cite{Milne}.
Notably, Sarkar and collaborators have argued that current Type Ia supernova (SN1a) data do not provide conclusive evidence for cosmic acceleration \cite{NGS, Sarkar2019}, while Blanchard et al. \cite{Tutu2017} have shown that acceleration inferred from ``local" cosmological probes, specifically those within the redshift range $z < 3$, lacks robust statistical significance. 
A comprehensive review of coasting cosmologies has been recently presented by Casado \cite{Casado2020}.

A coasting cosmology can be constructed either by positing some ad-hoc equation of state \cite{Melia2012}, or by assuming that the total energy-matter content of the universe is zero, as in the Milne cosmology \cite{Milne}. The latter case may be achieved if the the universe contains equal amounts of positive and negative mass. 
The existence of particles with negative mass has been a subject of theoretical speculation for decades, beginning with the seminal work of Hermann Bondi \cite{Bondi}. Although often dismissed as unphysical, negative mass is not inherently incompatible with foundational principles such as the equivalence principle, conservation laws, or Newton's third law \cite{Price_ajp1993, Mannheim_foundphys2000, Hammond_ejp2015}. Nonetheless, it introduces highly non-intuitive dynamics, such as the runaway acceleration of pairs of positive and negative masses \cite{Bonnor1989}, and peculiar gravitational behavior such as levitation and polarization in bound systems \cite{Price_ajp1993, Hammond_ejp2015}.

The concept of negative mass has been explored in various theoretical frameworks as a means to address outstanding problems in cosmology. Models incorporating negative mass components have been proposed to explain phenomena such as cosmic acceleration, dark energy, and the large-scale structure of the universe \cite{Petit2014,Farnes2018}. In particular, negative mass theories offer alternative mechanisms for repulsive gravitational effects without invoking a cosmological constant \cite{Farnes2018}. 
In the same vein, the so-called ``Dirac-Milne" universe is an alternative cosmological model that postulates a symmetric distribution of matter and antimatter, with antimatter possessing negative active gravitational mass \cite{BenoitLevy,Chardin2018}. Cosmological structure formation within this model was investigated recently, with intriguing results \cite{Manfredi2018,Manfredi2020}. However, recent measurements appear to rule out the hypothesis of repulsive gravity for antimatter \cite{Anderson2023}, although they cannot exclude a possible asymmetry between the gravitational responses of matter and antimatter (see, for instance, \cite{Menary2024,Menary2025}).

In the present work, we first introduce a new cosmological coincidence: a transition in the dynamical properties of the expanding universe that appears to occur precisely around the present epoch, like the usual cosmological coincidence.
This new coincidence involves the universe transitioning from a weakly coupled (collisionless) to a strongly coupled (collisional) gravitational regime. 
We then explore the possibility of a cosmological model in which the universe contains equal amounts of positive and negative mass, the latter understood in the sense proposed by Bondi \cite{Bondi}, which is fully compatible with the weak equivalence principle and General Relativity (GR). In this framework, Bondi-type negative masses possess negative inertial as well as negative active and passive gravitational mass.
Within such a “Bondi universe,” we demonstrate that the transition in the gravitational coupling coincides with the shift from a coasting to an accelerating expansion. This provides a unified reinterpretation of both coincidences as manifestations of the same underlying physical mechanism.

Numerical simulations of an exact one-dimensional (1D) model of Newtonian gravity, incorporating both positive and negative masses, support these results and reveal distinct expansion phases consistent with the proposed Bondi scenario. 
Although 1D gravity is, by construction, an idealized system, it has long been recognized as a valuable tool for probing the fundamental behaviour of self-gravitating systems \cite{Miller2023}. In addition, the runaway acceleration effect,  which is the key element of our Bondi universe, may not depend too strongly on the dimensionality of the system. 

However, one should remain aware of the substantial differences between 1D and 3D gravity, in particular because in 1D all particles move on the same line and can thus collide easily, unlike in 3D. Further, the mechanism by which Bondi pairs of positive/negative form and subsequently undergo runaway acceleration could be altered in 3D, where the Newtonian force weakens with the inverse square distance, instead of being uniform as in 1D. Finally, a strictly 1D framework suppresses angular momentum, thereby omitting an essential ingredient of realistic gravitational dynamics. All in all, although the runaway process of positive/negative mass pairs is likely to persist in 3D, the actual motion of the pairs may become far richer, potentially displaying more intricate orbital patterns and heightened sensitivity to perturbations.

This work is organized as follows. In Section \ref{sec:coincidence} we spell out the features of the coincidence involving the transition from a collisionless to a collisional universe, and show that such transition occurs around the present epoch. Section \ref{sec:negative} reviews the concept of negative mass in Newtonian gravity, particularly the negative mass concept put forward by Bondi \cite{Bondi}. In Section \ref{sec:linear}, we compute the linear response of a system containing equal amounts of positive and negative masses, in the weakly coupled regime described by a set of Vlasov-Poisson equations. We show that this system is always linearly unstable, irrespective of the temperatures and masses of the two species, and that the growth rate of the unstable modes is inversely proportional to the square root of the wavelength. Section \ref{sec:acceleration} contains numerical simulations of an N-body model with positive and negative Bondi masses. In particular, we show that the system undergoes both transitions --- from weakly to strongly coupled and from coasting to accelerating --- around the same epoch and because of the same mechanism.
Finally, in Section \ref{sec:conclusion}, we discuss the obtained results and their impact on our understanding of the universe expansion.


\section{Coincidence}\label{sec:coincidence}

Let us consider a gas of particles (``galaxies") of mass $m_g$, interacting through Newton's gravitational force. In analogy with a charged particle gas (plasma) \cite{Rouet2005,Gravier2023,Conde2018}, the coupling parameter $\Gamma$ for such a system may be written as the ratio of the typical gravitational potential energy to the typical kinetic energy:
\be
\Gamma = \frac{E_{\rm grav}}{E_{\rm kin}} = \frac{m_g^2 G d^{-1}}{k_B T},
\label{eq:gamma}
\ee
where $G$ is Newton's constant, $d$ is the mean distance between galaxies, $k_B$ is Boltzmann's constant, and $T$ is the temperature of the gas.
When the coupling parameter is small, $\Gamma \ll 1$, the gravitational gas is weakly coupled, and the mean gravitational field dominates over two-body correlations (i.e., collisions). In contrast, when $\Gamma \gtrsim 1$, the system is strongly coupled and collisions dominate over the mean field.

As in an ordinary gas, the temperature is related to the velocity dispersion around the mean velocity of the gas, so that we write: $k_B T = m_g v_p^2$, where $v_p$ is the root-mean-square velocity. In a cosmological scenario, the mean velocity is given by the Hubble flow, while the velocity dispersion is related to the so-called peculiar velocities $v_p$.
Hence, we can rewrite Eq. \eqref{eq:gamma} as
\be 
\Gamma_0 = \frac{m_g G }{d_0 v_{p0}^2},
\label{eq:gamma2}
\ee
where we have introduced the subscript ``0" to denote quantities calculated at the present epoch.
We consider typical values for the universe at the current epoch: intergalactic distance $d = 10^{22}\,\rm m \approx 10^6 \rm \,ly$ \cite{Karachentsev2005} and typical mass of a galaxy $m_g = 3 \times 10^{42} \,\rm kg \approx 1.5 \times 10^{12} \, M_\odot$ (estimated mass of the Milky Way \cite{Phelps2013}).
Using these values, we find $\Gamma_0 \approx 0.5$, which is remarkably close to unity.
Naturally, this estimate should be interpreted with caution, given the inherent uncertainties in the astrophysical data. Nevertheless, the result suggests that our universe lies near the threshold between a weakly coupled and a strongly coupled gravitational system.

In order to estimate the evolution in time of the coupling parameter $\Gamma(t)$, one needs an evolution law for $T(t)$, i.e., a gravitational equation of state. We take the simplest case, an adiabatic nonrelativistic relation of the type 
\[
\frac{T}{T_0} =  \left(\frac{\rho}{\rho_0} \right)^{\gamma -1},
\]
where $\rho$ is the mass density and $\gamma = 5/3$ is the nonrelativistic adiabatic exponent. In addition, the intergalactic distance goes as: $d(t) = d_0\, a(t)$, where $a(t)$ is the universe scale factor. As $\rho \thicksim a^{-3}$, we obtain that the temperature decays as the inverse of the squared scale factor: $ T(t) =  T_0 \,a(t)^{-2}$. For the coupling parameter, one gets $ \Gamma(t) =  \Gamma_0 \,a(t)$, suggesting that the universe was weakly coupled before our epoch and is currently transitioning to a strongly coupled regime. Note that such temperature dependence implies  $v_p \thicksim a^{-1}$, in accordance with the nonrelativistic expression \cite{Peacock2003}, i.e., peculiar velocities of nonrelativistic objects suffer redshifting by exactly the same factor as photon momenta. 

From the preceding results, it appears that there are actually {\em two} cosmological transitions that occur around the present epoch: (i) the usual transition from a decelerating (or coasting) expansion to an accelerating one, and (ii) the transition from a (gravitationally) weakly coupled regime to a strongly coupled regime. In order to explain this ``double coincidence problem", we will postulate a universe that contains an equal amount of positive and negative masses, in the Bondi sense \cite{Bondi}. Numerical simulations of this model reveal that the two transitions occur at the same time and are due to the same underlying mechanism.

The fact that, around the present epoch, the kinetic and gravitational potential energies are of the same order of magnitude is reminiscent of the well-known virial theorem in stellar dynamics (see Appendix~\ref{app:virial}). The virial theorem states that, at equilibrium, the following relationship should hold: $2K+U=0$, where $K$ and $U$ are, respectively, the kinetic and potential energies per unit mass.
This may appear similar to our above observation that $E_{\rm grav} \approx E_{\rm kin}$ at the current epoch of the universe expansion.

However, there are two important differences. First, the quantities involved are not the same. In the virial theorem formulation, $U$ and $K$ refer to the {\em total} energies of a closed self-gravitating system, such as a galaxy or a cluster of galaxies. In contrast, in the derivation of Eqs. \eqref{eq:gamma} and \eqref{eq:gamma2}, 
$E_{\rm grav}$ and $E_{\rm kin}$ represent {\em local} energies in a region of space that rides on the expanding background; in particular, $E_{\rm kin}$ is the peculiar kinetic energy (residual kinetic energy after subtracting the Hubble flow) and not the total kinetic energy.
Second, the virial theorem holds for a confined self-gravitating system at steady state, but not for the whole universe, which is not stationary because of the cosmological expansion. 
An extension of the virial theorem taking into account the expansion is provided by the Layzer-Irvine equation \cite{peebles2,Avelino2012}. But the latter is derived using comoving co-ordinates and assumes an expanding universe, while our analysis (see Sec. \ref{sec:acceleration}) is performed in fixed co-ordinates without {\it a priori} assuming any expansion.


\section{Negative mass}\label{sec:negative}

The possible existence of matter endowed with negative mass is a long-standing question that has received many, often contradictory, answers \cite{Price_ajp1993,Hammond_ejp2015,Bonnor1989,Nieto1991,Chardin1992,Chardin2018}. In this work, we will adopt the viewpoint first put forward by Hermann Bondi \cite{Bondi}, which remains, at least in principle, compatible with GR. Further, we consider a negative‑mass component that couples to ordinary matter solely through gravity, following an approach similar to that of Farnes \cite{Farnes2018}.
Such a construction represents a minimal extension of standard gravity: it leaves the well‑tested physics of the early universe, including primordial nucleosynthesis, mostly unaffected, while modifying only the subsequent, gravity‑dominated stages of cosmic evolution (after recombination), which are those relevant to the present work. Moreover, because its interactions are restricted to the gravitational sector, the negative-mass component should not interfere with the Standard Model of particle physics. 
In particular, our model is purely classical and therefore leaves aside the production of pairs of positive and negative mass particles that may occur in a quantum field theory of gravity. In today's absence of such a theory, it is not unreasonable to assume that  pair formation does not occur. We also emphasize that, in our approach, negative mass is not associated with antimatter, in contrast to the Dirac-Milne model \cite{BenoitLevy,Manfredi2018,Manfredi2020}.

Overall, our framework remains close in spirit to the \LCDM scenario --- where the dark energy component acts as a fluid with positive energy density and negative pressure --- but with one crucial distinction: in our case, 
the expansion is driven by a negative-mass component whose repulsive gravitational effect leads to runaway acceleration.
In both pictures, the dark-energy or negative-mass sector couples to luminous matter exclusively  through gravity. 
However, in the \LCDM model, dark energy (in the form of a cosmological constant) possesses a time-independent and spatially uniform density, whereas our negative-mass component exhibits a nontrivial gravitational dynamics.

In a Newtonian framework, we may distinguish three types of mass: the active gravitational mass $m_a$, the passive gravitational mass $m_p$, and the inertial mass $m_i$ \cite{Manfredi2018}. The active gravitational mass appears on the right-hand side of Poisson’s equation, which governs the gravitational potential $\phi$:
\be
\Delta \phi = 4\pi G \rho =  4\pi G m_a n,
\ee
where $n$ denotes the number density and $\rho$ the mass density. The passive gravitational mass determines the response of a body to a gravitational field, as expressed by the relation ${\bm F} = -m_p \nabla\phi$, where ${\bm F}$ is the force. Finally, the inertial mass characterizes the relationship between momentum and velocity, given by ${\bm p} = m_i \dot {\bm r}$, where the dot represents differentiation with respect to time. Consequently, Newton’s second law, $\dot {\bm p}={\bm F}$, yields the equation of motion: $ \ddot {\bm r} = -(m_p/m_i) \nabla\phi$.

The equivalence principle (EP) asserts the identity of inertial and passive gravitational masses, $m_i = m_p$, thereby rendering a gravitational field locally indistinguishable from an acceleration. However, the EP imposes no constraint on the active gravitational mass, which may, in principle, differ from the other two.
In contrast, Newton’s third law --- the principle of action and reaction --- requires equality between the active and passive gravitational masses, $m_a = m_p$, to ensure that the net force on an isolated system of interacting particles vanishes, thereby preserving momentum conservation.
Combining these two fundamental principles leads to the conclusion that all three masses must be equal: $m_a = m_p = m_i$ \footnote{For an overview of Newtonian and non-Newtonian scenarios with negative mass, see our earlier work \cite{Manfredi2018}}. However, this relationship is respected not only for the usual case where all masses are positive, but also for the unconventional scenario where they are all negative. Such configuration corresponds to the so-called Bondi masses \cite{Bondi}, and can be summarized as follows: positive masses attract all other masses, while negative masses repel all others. As shown by Bondi, this scenario is compatible with GR, as both positive and negative masses follow the same geodesics.

This framework, as noted by several authors \cite{Price_ajp1993,Hammond_ejp2015,Bonnor1989}, gives rise to surprising dynamical phenomena, such as the runaway acceleration of a positive/negative mass pair. Indeed, if a positive and a negative mass are initially at rest and placed in proximity, the positive mass attracts the negative one, while the negative mass repels the positive. The result is a self-sustaining uniform acceleration of both masses, while the total momentum of the system remains conserved (i.e., zero). 

For our present purposes, the existence of the runaway effect is a crucial feature, because it can naturally lead from a coasting expansion ($a \thicksim t$) to a uniformly accelerating expansion ($a \thicksim t^2$).  As the runaway acceleration is a two-body effect, and not a mean-field effect, it is also natural that the acceleration becomes relevant only when the self-gravitating system goes from a weakly coupled to a strongly coupled regime.
The precise mechanism will be detailed in the following sections.

Such accelerating mechanism is reminiscent of the backreaction effect, which has been proposed as a possible alternative to dark energy in a cosmological setting \cite{Schander2021,Green2014,Buchert2015}. 
Backreaction refers to the idea that small-scale inhomogeneities -- such as galaxies, clusters, and voids -- can influence the large-scale dynamics of the universe. While standard cosmological models treat the universe as homogeneous and isotropic on large scales, backreaction explores how the nonlinear evolution of these structures might affect cosmic expansion, challenging the assumption that averaging over local variations yields the same behavior as a perfectly uniform universe.
Intriguingly, cosmological backreaction also stems from a coincidence: the universe's accelerated expansion begins at the very epoch when large-scale gravitational structures have fully formed. This temporal alignment suggests a possible link between the nonlinear growth of cosmic inhomogeneities and the onset of acceleration.

However, the principal idea put forward in the present work is fundamentally different. Indeed, structure formation, and hence inhomogeneities, can well arise in a weakly coupled (collisionless) model of the expanding universe---there is no need for two-body correlations to play a significant role. For instance, in an earlier work \cite{Manfredi_PRE2016}, the present authors explored the formation and evolution of gravitational structures using a purely mean-field model (Vlasov-Poisson equations).
Here, in contrast, we investigate a scenario where the acceleration is due to the universe transitioning from a weakly coupled to a strongly coupled regime, where two-body correlations lead to the Bondi runaway acceleration effect, and ultimately to a globally accelerating expansion.

Cosmological simulations using negative Bondi masses were reported in a recent study by Farnes \cite{Farnes2018}, although some of the results were later criticized \cite{SocasNavarro2019}. An unpleasant feature of Farnes' model is that it has to resort to negative mass creation in order to mimic a cosmological constant, as the energy density of matter scales as $a^{-3}$, while the energy density associated with the cosmological constant (replaced, in his work, by a negative mass fluid) should remain constant over time. 
On the other hand, Farnes \cite{Farnes2018} reports no observation of runaway pairs in his runs, which may be attributed to the relatively modest number of particles used ($N = 50{,}000$) for a 3D simulation. Our own simulations use a similar number of particles, but are restricted to one spatial dimension, thus guaranteeing a sufficient sampling of the relevant phase-space volume.

In addition, as is illustrated in Section \ref{sec:acceleration}, it is necessary to follow the evolution for a very long time before one can observe the accelerated expansion. Unlike the work of Farnes \cite{Farnes2018}, the N-body code adopted here is ``exact", in the sense that it integrates the equations of motion with no numerical errors other than those due to the finite representation of real numbers on the computer \cite{Miller2023}. Hence, very accurate results can be obtained over extremely long simulation times, enough to observe the transition to the uniformly accelerated regime.


\section{Weakly-coupled systems: linear response}\label{sec:linear}

Before performing numerical simulations of the full N-body dynamics, which show the transition from the weakly coupled to the strongly coupled regimes, we report here some analytical results for weakly coupled systems.

The weakly coupled (collisionless) dynamics of a gravitational system made of Bondi masses may be described, in the Newtonian limit, by the following Vlasov-Poisson equations \cite{Manfredi2018}:
\begin{eqnarray}
\frac{\partial f_\pm}{\partial t} &+&
{\bm v}\cdot \nabla f_\pm + {\bm E}  \cdot \nabla_{\bm v} f_\pm = 0, \label{eq:vlasov} \\
\nabla \cdot {\bm E} &=& -4\pi G  \left(m_+ n_{+} + m_- n_{-}\right), \label{eq:poisson}
\end{eqnarray}
where $f_{\pm}({\bm r},{\bm v},t)$ are the probability densities (distribution functions) in the phase space for positive and negative mass particles,  $n_{\pm}({\bm r},t)= \int f_{\pm} d{\bm v}$ are the number densities,  ${\bm E} = -\nabla \phi({\bm r},t)$ is the gravitational field, and $\phi({\bm r},t)$ the gravitational potential. The total net mass is assumed to be zero: $\int \left(m_+ n_{+} + m_- n_{-}\right) d{\bm r} =0$.
Note that the Vlasov equations \eqref{eq:vlasov} are identical for positive and negative masses: this is the Newtonian signature that both species follow the same geodesics in GR.

In this section, we study the linear response of Eqs. \eqref{eq:vlasov}-\eqref{eq:poisson}, following the same procedure as in \cite{Manfredi2018}. We expand all quantities around a spatially uniform equilibrim denoted by the subscript ``eq": the distribution functions $f_\pm({\bm r},{\bm v},t)=f_{eq \pm}({\bm v}) + \tilde f_{\pm}({\bm r},{\bm v},t)$,  the densities $n_\pm=n_{eq\pm} + \tilde n_{\pm}({\bm r},t)$, and the potential $\phi({\bm r},t) = \phi_{eq}+\tilde\phi({\bm r},t)$ (where the tilde denotes a small perturbation) and only retain first-order terms.
For a homogeneous equilibrium, one has $m_{+} n_{eq+} + m_{-} n_{eq-}=0$, and the gravitational potential vanishes: $\phi_{eq}=0$.
Then, we Fourier analyze the perturbations in space and time, by writing for the density fluctuations:
$\tilde n_{\pm}({\bm r},t) = \tilde n_{\pm}\,\exp[i({\bm k}\cdot{\bm r}-\omega t)]$,
and similarly for the other first-order quantities.

In line with the 1D numerical simulations presented in Sec. \ref{sec:acceleration}, we consider here 1D perturbations to the Eqs. \eqref{eq:vlasov}-\eqref{eq:poisson}. 
Hence, in the following, $E$, $k$ and $v$ will refer respectively to the gravitational field, wave vector, and velocity along the direction of propagation of the linear waves.
By so doing, we obtain the linear relations:
\begin{eqnarray}
\tilde f_{\pm} &=& -i \tilde E \,\frac{\partial_v f_{eq \pm}}{\omega - k v}, \label{eq:vlasov-lin} \\
i k \tilde E &=& -4\pi G \int_{-\infty}^{\infty} \left(m_+ \tilde f_{+} + m_- \tilde f_{-}\right) dv . \label{eq:poisson-lin}
\end{eqnarray}
Substituting the Eqs. \eqref{eq:vlasov-lin} into Eq. \eqref{eq:poisson-lin}, we get
\begin{equation}
k= 4\pi G \int_{-\infty}^{\infty} \left(m_+ \,\frac{\partial_v f_{eq +}}{\omega - k v} + m_- \,\frac{\partial_v f_{eq -}}{\omega - k v} \right) dv. 
\end{equation}
Finally, expanding the denominator in the limit $kv \ll \omega$,
\[
\frac{1}{\omega - k v} = \frac{1}{\omega} \,\left(1 +\frac{kv}{\omega} +\frac{k^2v^2}{\omega^2} +\frac{k^3 v^3}{\omega^3} + \dots\right),
\]
and integrating by parts in velocity space, we are lead to the following dispersion relation, valid up to second order in $k$:
\be
\omega^4 = 3 \,\omega_J^2 \, k^2 \left(v_{T-}^2 -v_{T+}^2 \right),
\label{eq:disprel}
\ee
where $\omega_J = \sqrt{4\pi G |m_{\pm}| n_{eq\pm}}$ is the equilibrium Jeans frequency (the same for both species), and
$v_{T \pm}^2 = \int f_{eq \pm}(v)\, v^2 dv = k_B T_{\pm}/m_{\pm}$, where the last equality holds for a Maxwell-Boltzmann equilibrium.
The above dispersion relation is valid in the limit of small $k$ (long wavelengths) and neglects Landau damping, which would have required the full treatment of the pole originating from the denominator $\omega - k v$ in Eq. \eqref{eq:vlasov-lin}.

Equation \eqref{eq:disprel}, being of fourth order in $\omega$, possesses four complex solutions, one of which displays a positive imaginary part. Hence, an unstable solution always exists, regardless of the sign of the right-hand side, except in the special case of equal thermal velocities, $v_{T-} = v_{T+}$ \footnote{This was the case analyzed in Ref. \cite{Manfredi2018}, for which the linear dispersion relation is trivial, and it is necessary to go to second order to observe a nontrivial response.}.
The growth rate of the unstable modes is proportional to the square root of the wave number, i.e. ${\rm Im} (\omega) \sim \sqrt{k}$, hence shorter wavelengths grow more rapidly. This is very different from the usual Jeans instability, where only wavelengths longer than $v_T/\omega_J$ are unstable. 
Some numerical simulations with periodic boundaries highlighting this instability are shown in the Appendix \ref{app:periodic}.

When wavelengths close to the interparticle distance $d$ become dominant, the assumptions underlying the collisionless Vlasov-Poisson model cease to hold, and the system undergoes a transition to a strongly coupled (collisional) regime. Thus, by favoring short wavelengths, the dispersion relation \eqref{eq:disprel} may represent a plausible framework for understanding the transition from a weakly coupled to a strongly coupled universe. The numerical simulations reported in the next section will provide further details on this transition and its relation to the usual cosmological coincidence problem.


\section{Simulations of cosmic acceleration}\label{sec:acceleration}

In this section, we present simulation results obtained with a 1D N-body code that tracks the Newtonian dynamics of equal numbers of particles with positive and negative Bondi masses. As mentioned earlier, the code is exact because it is event-driven: it integrates exactly the dynamics of the 1D particles in between two particle crossings. Hence, it introduces no truncation errors associated with finite time steps or grid discretization; the only numerical errors arise from the finite precision of floating-point arithmetic. This property enables us to follow the system’s evolution over very long times with high accuracy. A brief description of the N-body algorithm is provided in our earlier review \cite{Miller2023}. 

It is worth emphasizing that our simulations are performed in fixed (not comoving) co-ordinates and that no {\em a priori} assumption on the expansion rate is made. In essence, we simulate a 1D self-gravitating system with open boundaries (the ``universe"), starting from a given initial condition at $t=0$. The expansion factor $a(t)$ is then obtained from the simulations as the root-mean-square displacement of such self-gravitating system: $a(t) = \sigma_{X}(t) \equiv \sqrt{\langle x^2 \rangle}$, where the average is computed over all the particles, as explained below.

The following simulations use $N = N_{+} + N_{-} = 2 \times 10^4$ particles of identical mass in absolute value, with $N_{+} = N_{-}$. The initial distribution is uniform in space between $-L/2$ and $L/2$, and Maxwellian in velocity, with initial velocity dispersion $\sigma_{V}(0)$.
We shall use units such that time is normalized to the inverse Jeans frequency $\omega_J$, space to the initial inter-particle distance $d$, densities to the total initial density, and velocities to $d \omega_J$. With this normalization, the initial extension of the system is simply $L = N$.

\subsection{Mean-square displacement and velocity}\label{sec:acceleration-A}
We shall focus on the expansion law, expressed through the mean square displacement $\sigma_{X}^{2}(t) = \langle x^2 \rangle = {1 \over N} \sum_{j=1}^{N} x_j^2$, where the $x_j$ are the positions of the $N$ particles and we have assumed that their average position is negligible, i.e. $\langle x \rangle \approx 0$.
The root-mean-square displacement $\sigma_{X}(t)$ is a good ersatz for the scale factor $a(t)$.

Before showing the numerical results, we illustrate how such a two-species self-gravitating system should go through three different phases of expansion, assuming that it starts in an almost collisionless ($\Gamma \ll 1$) initial state:

\begin{itemize}
\item
When the coupling parameter is small, $\Gamma \ll 1$, the kinetic energy dominates over the gravitational energy, see Eq. \eqref{eq:gamma}. During this initial phase, the self-gravitating system behaves as a non-interacting gas, hence its spatial width increases linearly in time, $\sigma_{X}(t) \thicksim t$;
\item
Because of the expansion, the system cools down and consequently becomes more correlated. For intermediate values of the coupling parameter $\Gamma$, the self-gravitating gas is still dominated by the kinetic energy, but, occasionally, a positive and a negative mass particles come into contact and directly interact gravitationally. During such an encounter, the two particles experience a Bondi-type acceleration in the form of a ``kick" in their velocities, whose sign depends on the relative position of the positive and negative particles ($+/-$ or $-/+$). This sequence of virtually random kick events leads to a random walk in velocity space, with the root-mean-square velocity increasing as $\sigma_{V}(t) \equiv \sqrt{\langle v^2 \rangle} \thicksim t^{1/2}$ \footnote{The possibility of stochastic acceleration that mimics dark energy was proposed in two recent works \cite{Lapi2023,Lapi2025}.}. Hence, the root-mean-square displacement goes like $\sigma_{X}(t) \thicksim t^{3/2}$;
\item
Finally, in the strongly coupled regime, when $\Gamma \approx 1$, pairs of positive/negative mass particles are formed and remain stable. Because of Bondi's runaway mechanism, each pair accelerates uniformly, either to the left or to the right, depending on their relative position. Consequently, the root-mean-square displacement increases as $\sigma_{X}(t) \thicksim t^{2}$.
\end{itemize}

\begin{figure}
    \centering
    \includegraphics[width=0.65\linewidth]{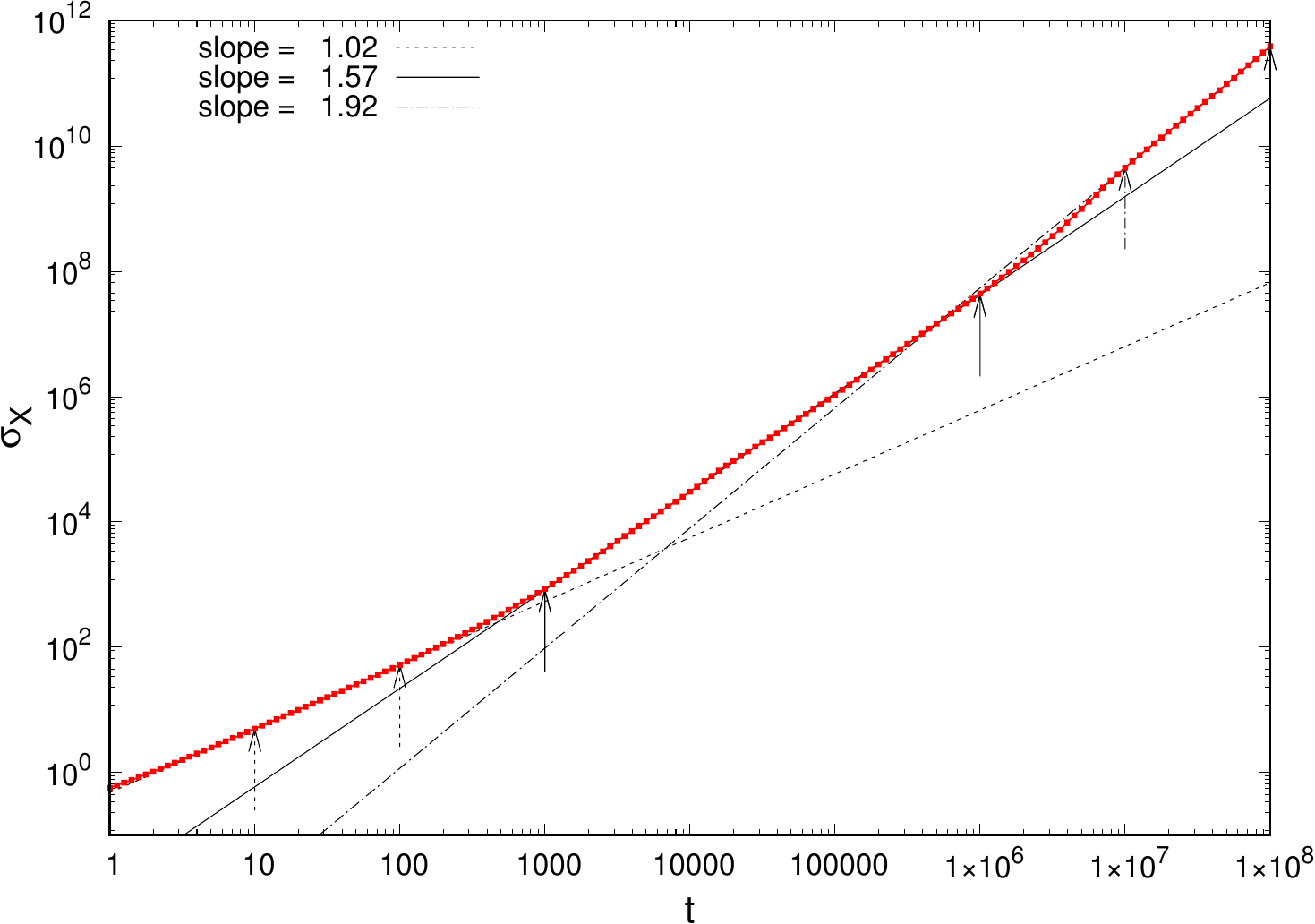}
    \caption{Time evolution, on a log-log scale, of the root-mean-square displacement $\sigma_{X}(t)$ for an initially collisionless system with velocity dispersion equal to $\sigma_{V}(0) = 10^4$ (red dots). The function is $\sigma_{X}(t)$ is fitted with three different power laws of the type $t^{\alpha}$ in the regions delimited by the arrows. The values of $\alpha$, computed via a least-square method, are found to be: $\alpha =1.02$ ($10<t<100$), $\alpha =1.57$ ($10^3<t<10^6$), and $\alpha =1.92$ ($10^7<t<10^8$).  
    The black straight lines represent the corresponding power laws. }
    \label{fig:sigma204}
\end{figure}

We stress that the mean-square velocity $\sigma_{V}^2 \equiv \langle v^2 \rangle$ is {\em not} representative of the peculiar velocities. Here, we do not use comoving co-ordinates, and therefore we do not split the velocity into a Hubble flow part and a peculiar part. 
\GM{Instead, $\langle v^2 \rangle$ represents (twice) the total kinetic energy per unit mass of the whole self-gravitating system, which is precisely the quantity $K$ that  enters the expression of the virial theorem, see Appendix \ref{app:virial}.
}

In Fig. \ref{fig:sigma204}, we show the time evolution of the root-mean-square displacement $\sigma_{X}(t)$, for an initially weakly-coupled system, where the initial root-mean-square velocity is $\sigma_{V}(0) = 10^4$. According to the definition of Eq. \eqref{eq:gamma2}, the coupling parameter can be written, in our units, $\Gamma = 1/\sigma_{V}^2$, and hence it is very small here. The figure shows that the root-mean-square displacement obeys a power law $\sigma_{X}(t) \thicksim t^{\alpha}$, with the exponent going from $\alpha=1$ to $\alpha=3/2$, and finally $\alpha=2$ for very long times, as expected from the above considerations.

\begin{figure}
    \centering
    \includegraphics[width=0.65\linewidth]{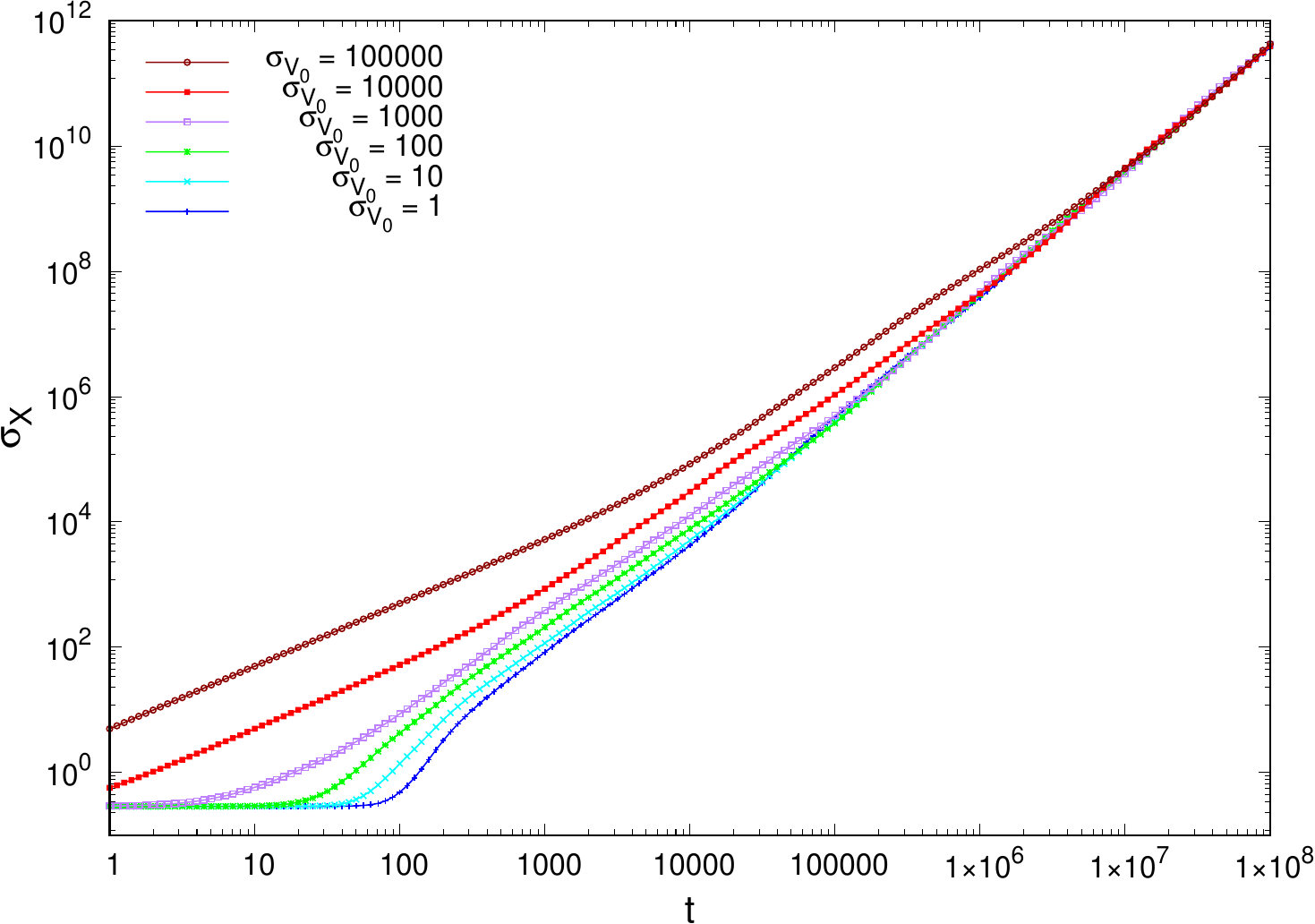}
    \caption{Time evolution, on a log-log scale, of the root-mean-square displacement $\sigma_{X}(t)$ for several simulations with initial velocity dispersions in the range $1 \le \sigma_{V}(0) \le 10^5$. Although the initial evolutions are different, they all converge to the power law $\sigma_{X}(t) \thicksim t^2$, for times $t \gtrsim 10^7$.}
    \label{fig:sigma20x}
\end{figure}

\begin{figure}
    \centering
    \includegraphics[width=0.65\linewidth]{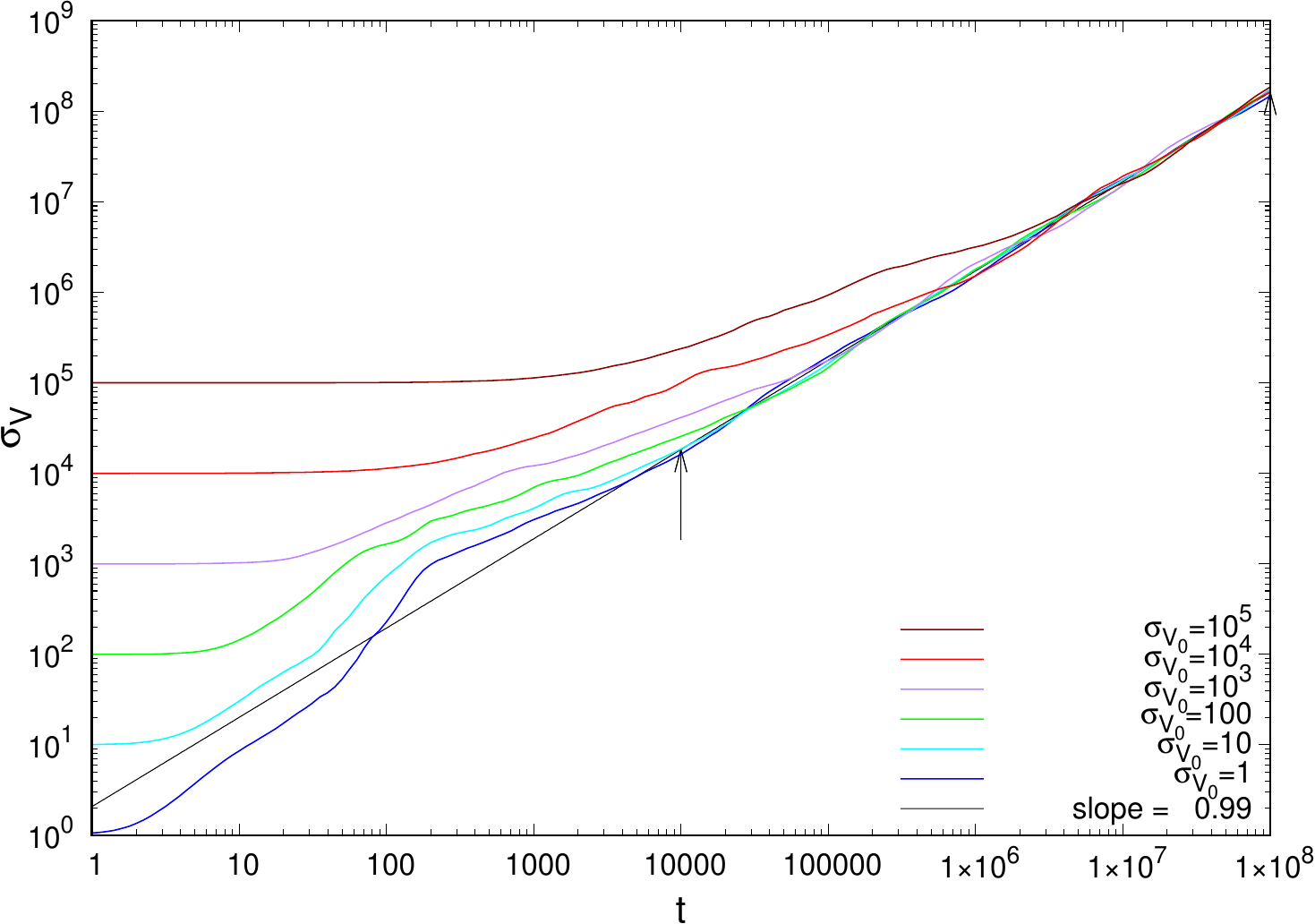}
    \caption{Time evolution, on a log-log scale, of the root-mean-square velocity  $\sigma_{V}(t)$ for several simulations with initial velocity dispersions in the range $1 \le \sigma_{V}(0) \le 10^5$. Although the initial evolutions are different, they all converge to the power law $\sigma_{V}(t) \thicksim t$, for times $t \gtrsim 10^7$. This power law is represented by the continuous black line.}
    \label{fig:sigma20v}
\end{figure}

Next, we show the same result for different values of the initial velocity dispersion, in the range $1 \le \sigma_{V}(0) \le 10^5$ (Fig. \ref{fig:sigma20x}), corresponding to different values of the initial coupling parameter. The evolutions are different --- in particular, we note a delay in the onset of the expansion for smaller values of $\sigma_{V}(0)$ --- but they all converge to the same power law $\sigma_{X}(t) \thicksim t^2$, for times $t > 10^7$. For the same cases, the plots of the root-mean-square velocity $\sigma_{V}(t) \equiv \sqrt{\langle v^2 \rangle(t)}$, shown in Fig. \ref{fig:sigma20v}, confirm these results: for long times, $\sigma_{V}(t)$ grows linearly in $t$. This is the expected law when Bondi pairs have formed and are stable. 

The initial delay observed in Fig. \ref{fig:sigma20x} can be explained by the dominating gravitational interactions for moderate values of $\Gamma$ [large $\sigma_{V}(0)$], which freeze the particle positions, until the residual kinetic energy has had enough time to break down this configuration by inducing some particle crossings. Indeed, for  $\Gamma \gg 1$, the system can be viewed as a solid  ``crystal", where the particles never cross each other but only slightly vibrate.


\subsection{Matter density and phase-space distributions}

\begin{figure}
    \centering
    \includegraphics[width=1\linewidth]{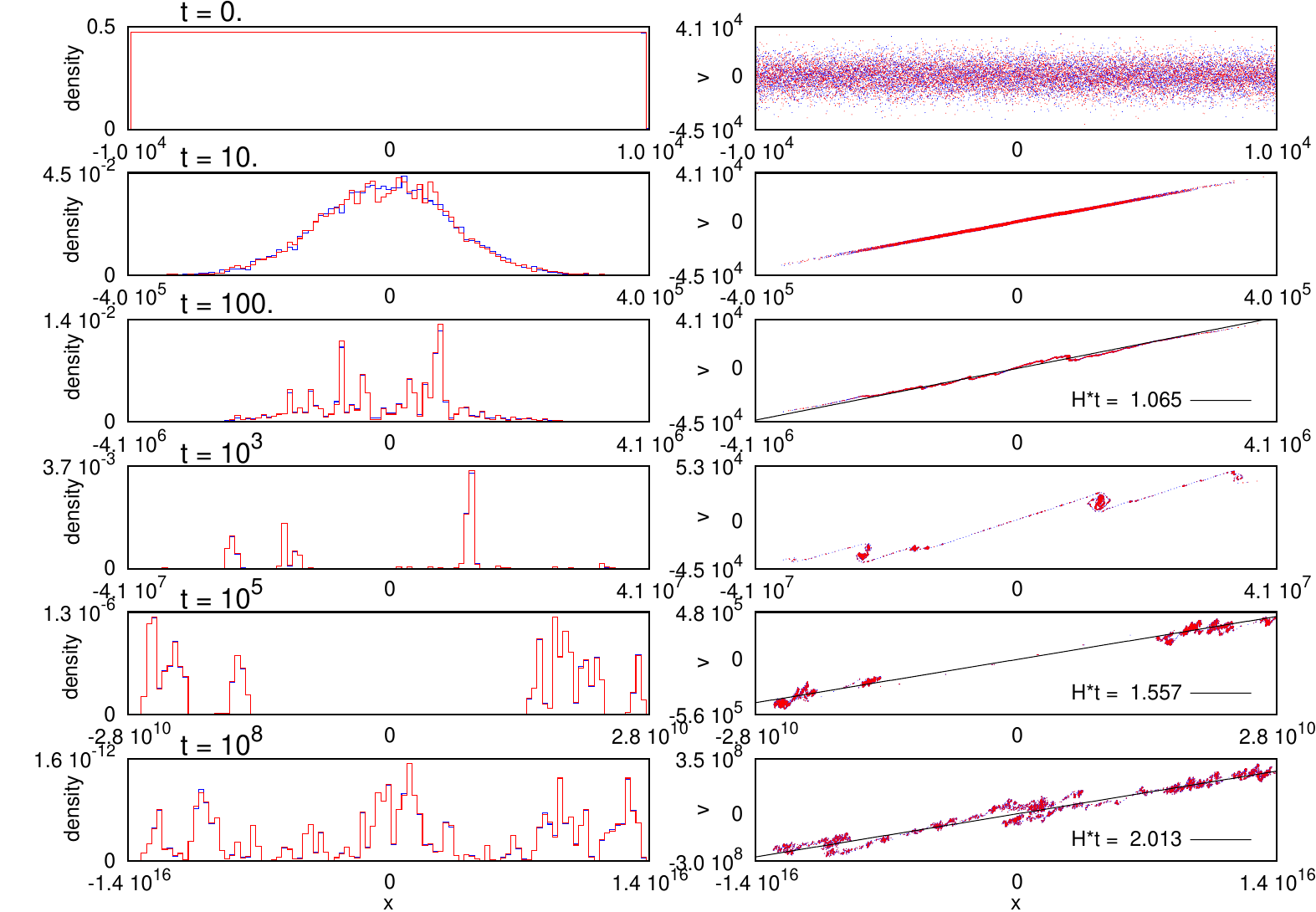} 
    \caption{Particle density (left panels) and phase space distributions (right panels) for an initially weakly correlated system with velocity dispersion $\sigma_{V}(0) = 10^4$. Positive mass particles are depicted in red, while negative mass particles are depicted in blue. The straight lines in the phase-space plots represent the slope $H= \alpha/ t$ of the filament (instantaneous Hubble parameter), where the product $H t$ is expected to approach the exponent $\alpha$ of the expansion law. Note that both the position and velocity scales change over time, in order to capture the expansion of the system.    
    }
    \label{fig:phasespace204}
\end{figure}

\begin{figure}
    \centering
    \includegraphics[width=1\linewidth]{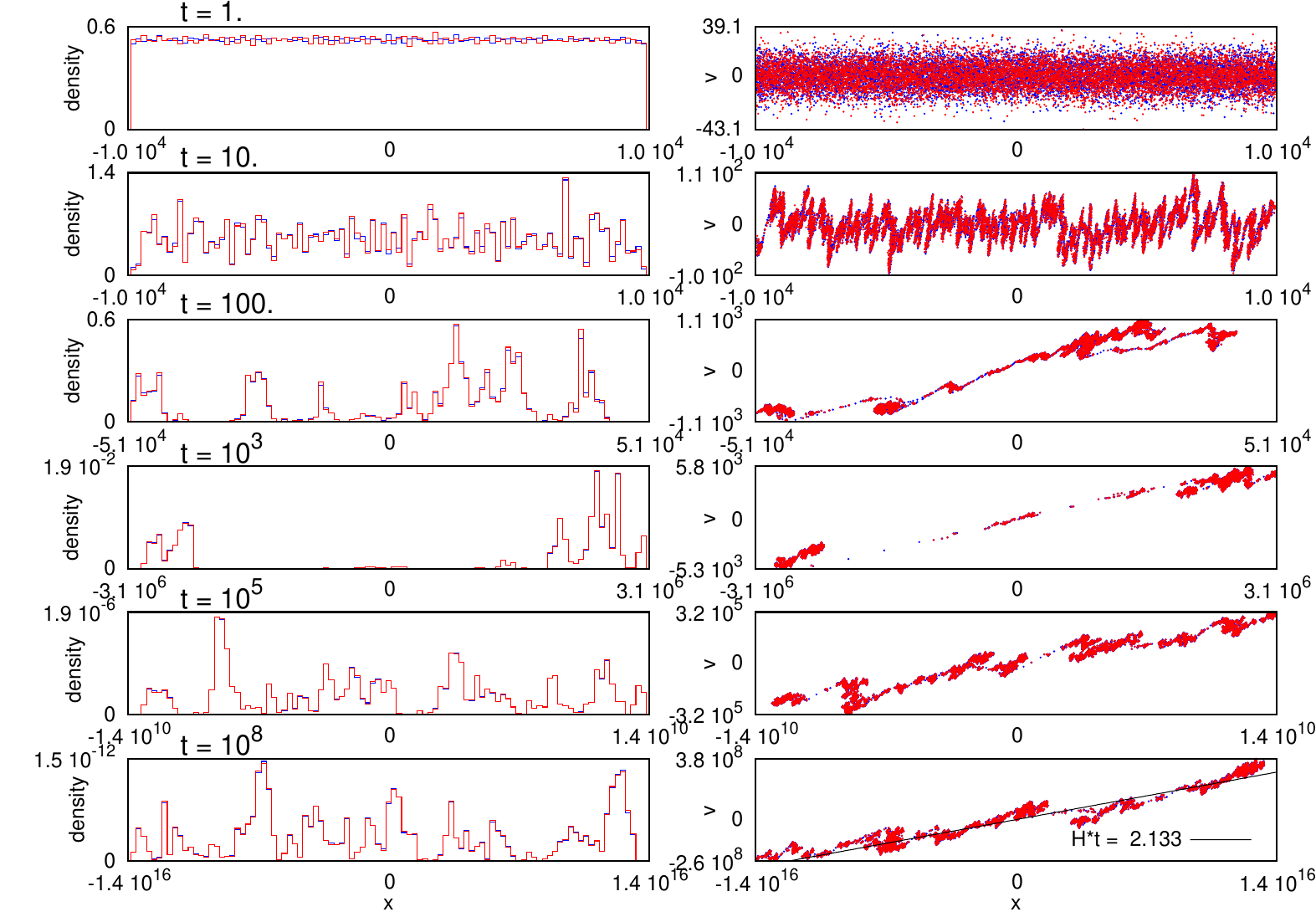} 
    \caption{Particle density (left panels) and phase space distributions (right panels) for an initially strongly correlated system with  velocity dispersion $\sigma_{V}(0) = 10$. Positive mass particles are depicted in red, while negative mass particles are depicted in blue. The straight line in the phase-space plot at $t=10^8$ represents the slope $H= \alpha/ t$ of the filament (instantaneous Hubble parameter), where the product $H t$ is expected to approach the exponent $\alpha$ of the expansion law. Note that both the position and velocity scales change over time, in order to capture the expansion of the system.}    
    \label{fig:phasespace201}
\end{figure}

For the same case of Fig. \ref{fig:sigma204}, we plot in Fig. \ref{fig:phasespace204} the particle density and the phase-space density at several instants in time. The latter plots are particularly interesting, as they bear the signature of the power-law expansion  $x \thicksim t^{\alpha}$. The velocity expansion law is then $v \thicksim \alpha\, t^{\alpha-1}$, so that one has: $v = {\alpha \over t} x = H x$. The relation $v = H x$ corresponds precisely to the Hubble law (speed proportional to distance), with a time-dependent Hubble parameter $H = \alpha/t$. This Hubble law emerges naturally within the framework of our model.
The slope of the filament-like structure in the right column of Fig. \ref{fig:phasespace204} is therefore expected to decrease with time in such a way that the product $H t$  matches the exponent $\alpha$. This is precisely what is observed in the simulation, for the three expansion phases that correspond to $\alpha = 1$ ($t=100$), $\alpha = 3/2$ ($t=10^5$), and $\alpha = 2$ ($t=10^8$).

We also observe the emergence of nonlinear gravitational structures, which sets in shortly after the conclusion of the initial ballistic phase ($t \approx 100$). These structures involve both positive and negative masses, consistent with the fact that all masses follow identical geodesics, i.e., the same Newtonian trajectories. First, several clusters of particles appear, connected by thin filaments ($t \approx 10^3$). At later times, these clusters progressively migrate toward the boundaries of the system, while the central region becomes depleted ($t \approx 10^{5}$); this stage corresponds to the establishment of the intermediate expansion regime with exponent $\alpha = 3/2$. Finally, matter redistributes itself throughout the entire occupied domain ($t \approx 10^{8}$), a process that appears to coincide with the onset of the accelerated phase characterized by $\alpha = 2$.

In Fig. \ref{fig:phasespace201}, we show the density and phase-space plots for a system that is initially strongly coupled [$\sigma_{V}(0) =10$]. We note that structure formation occurs slightly earlier here. The accumulation of clusters near the edges of the system is also observed earlier than in the preceding case, around  $t \approx 10^3$, followed by mass redistribution throughout the entire domain. At late times, the phase-space slope reaches the value $Ht = \alpha \approx 2$, as in the initially weakly coupled case of Fig. \ref{fig:phasespace204}.

Finally, in a short movie presented in the Appendix \ref{app:movie}, we show the phase-space dynamics for a relatively small system ($L=N =500$), which makes it easier to follow individual particles. The movie shows the emergence of stable Bondi pairs that accelerate uniformly. Occasionally, one pair changes its ``polarity", i.e., the relative positions of the positive and negative masses, and therefore reverses its direction of acceleration. The simulation stops when no more particle crossings are observed. From that time on, the system evolves as a gas of independent Bondi pairs, each moving with constant acceleration in either direction.


\subsection{Comoving co-ordinates and power spectra} \label{sec:comoving}
Although we used fixed space-time co-ordinates in the simulations, it may be helpful to switch to comoving co-ordinates in order to better analyze the gravitational structure formation. Comoving co-ordinates follow the expansion of a self-gravitating system according to its scale factor $a(t)$. Here, quite naturally, we choose the scale factor to be the root-mean-square displacement of the particles as obtained from the numerical simulations, i.e., $a(t) = \sigma_{X}(t)$. The comoving co-ordinates $(\hat x, \hat v)$ are defined in terms of the fixed co-ordinates $(x,v)$ according to the transformations:
\begin{eqnarray}
x &=& a(t) \,\hat x , \\
v &=& a(t)\, {\hat v} + {\dot a}(t)\, {\hat x},
\end{eqnarray}
where the overdot denotes differentiation with respect to $t$. 
By rewriting $v = a(t) {\hat v} + H x$, where $H=\dot a/a$ is the Hubble parameter, we see that $a \hat v$ represents the deviations from the Hubble flow, i.e., the peculiar velocity.

\begin{figure} 
    \centering
    \includegraphics[width=0.48\linewidth]{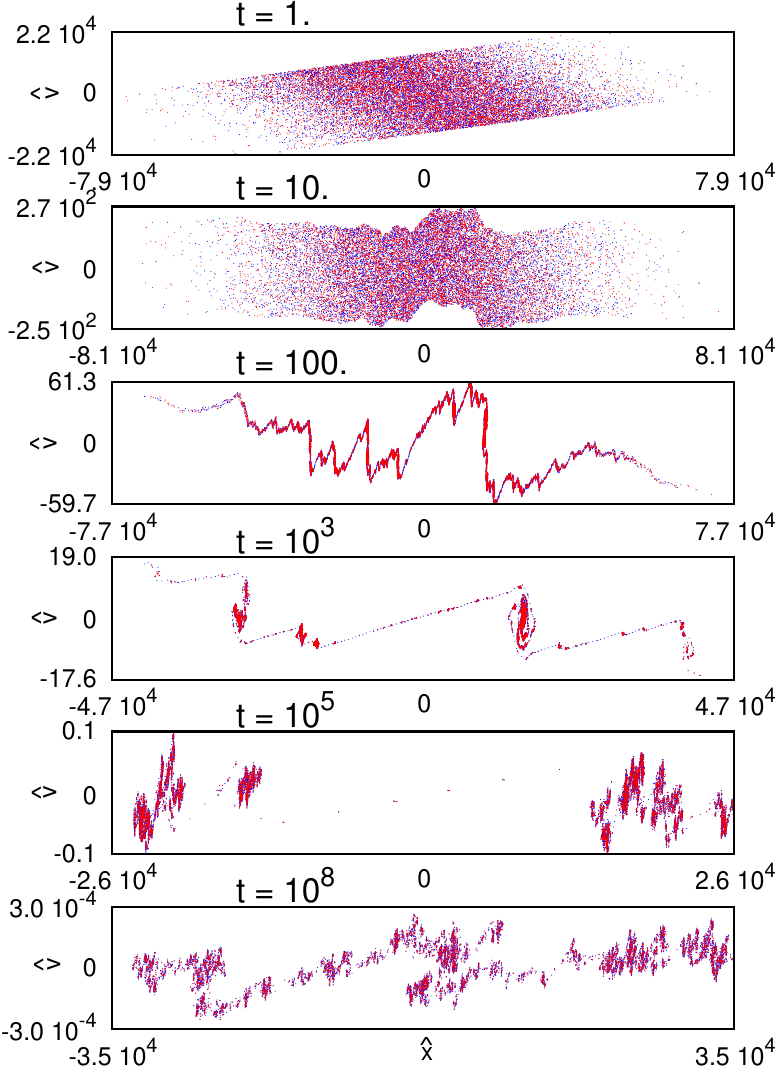} 
    \includegraphics[width=0.48\linewidth]{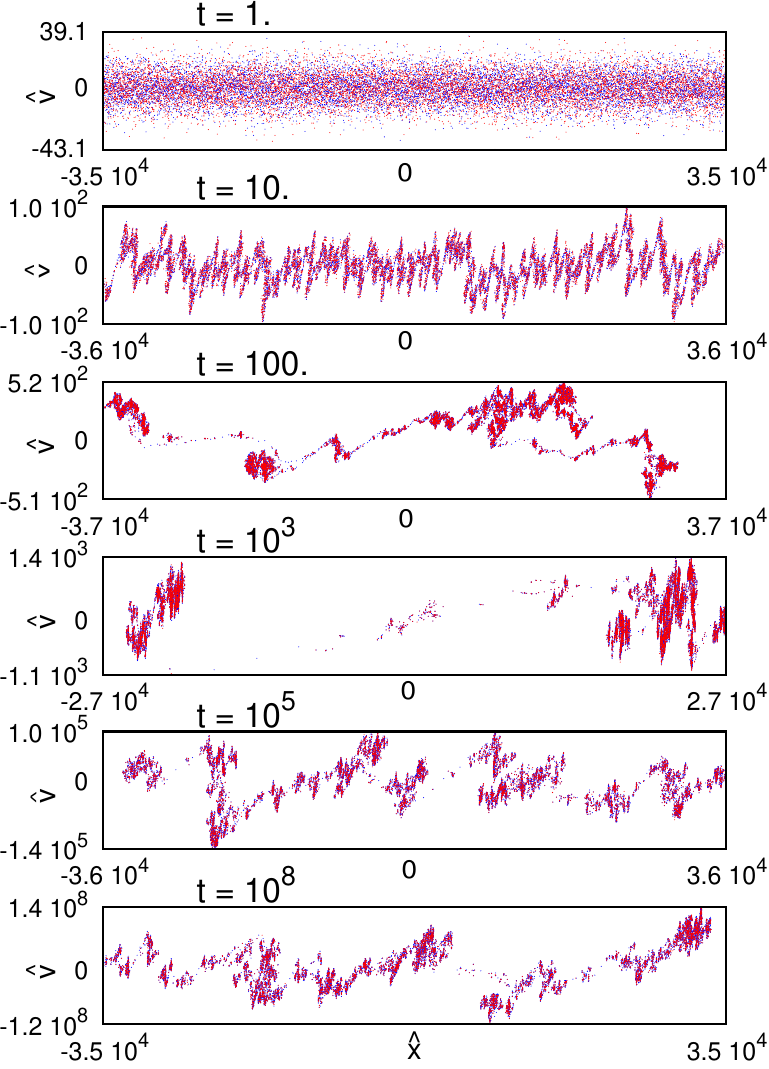} 
    \caption{Phase-space portraits of the particle distributions in the comoving co-ordinates $(\hat x, \hat v) $. Left panels: initially weakly coupled system with velocity dispersion $\sigma_{V}(0) = 10^4$. Right panels: initially strongly coupled system with $\sigma_{V}(0) = 10$. Note that the co-moving position $\hat x$ remains approximately the same during the entire evolution, signalling that the transformation has indeed absorbed the expansion factor.}
    \label{fig:phasespace-scaled}
\end{figure}

In Fig. \ref{fig:phasespace-scaled}, we show the phase space plots in comoving co-ordinates, for an initially weakly coupled case [$\sigma_{V}(0) = 10^4$, left panels] and an initially strongly coupled case [$\sigma_{V}(0) = 10$, right panels].
Note that the scaled spatial axis remains approximately within the same bounds for all times, signalling that the comoving co-ordinates absorb the expansion factor, as expected.

The representation in such scaled co-ordinates allows one to better visualize the formation of gravitational structures, with clusters and sub-clusters of particles. The structure formation begins earlier, near $t=100$, for the initially strongly coupled case (right panels), whereas the first structures only appear around $t=10^3$ for the weakly coupled case (left panels).
The depleted region at the center of the system is clearly visible at $t=10^5$ for the weakly coupled case and at  $t=10^3$ for the strongly coupled one.
For late times ($t \approx 10^8$), corresponding to the accelerated expansion $a(t)
 \sim t^2$, both phase-space distributions look very much alike.

\begin{figure} 
    \centering
    \includegraphics[width=0.48\linewidth]{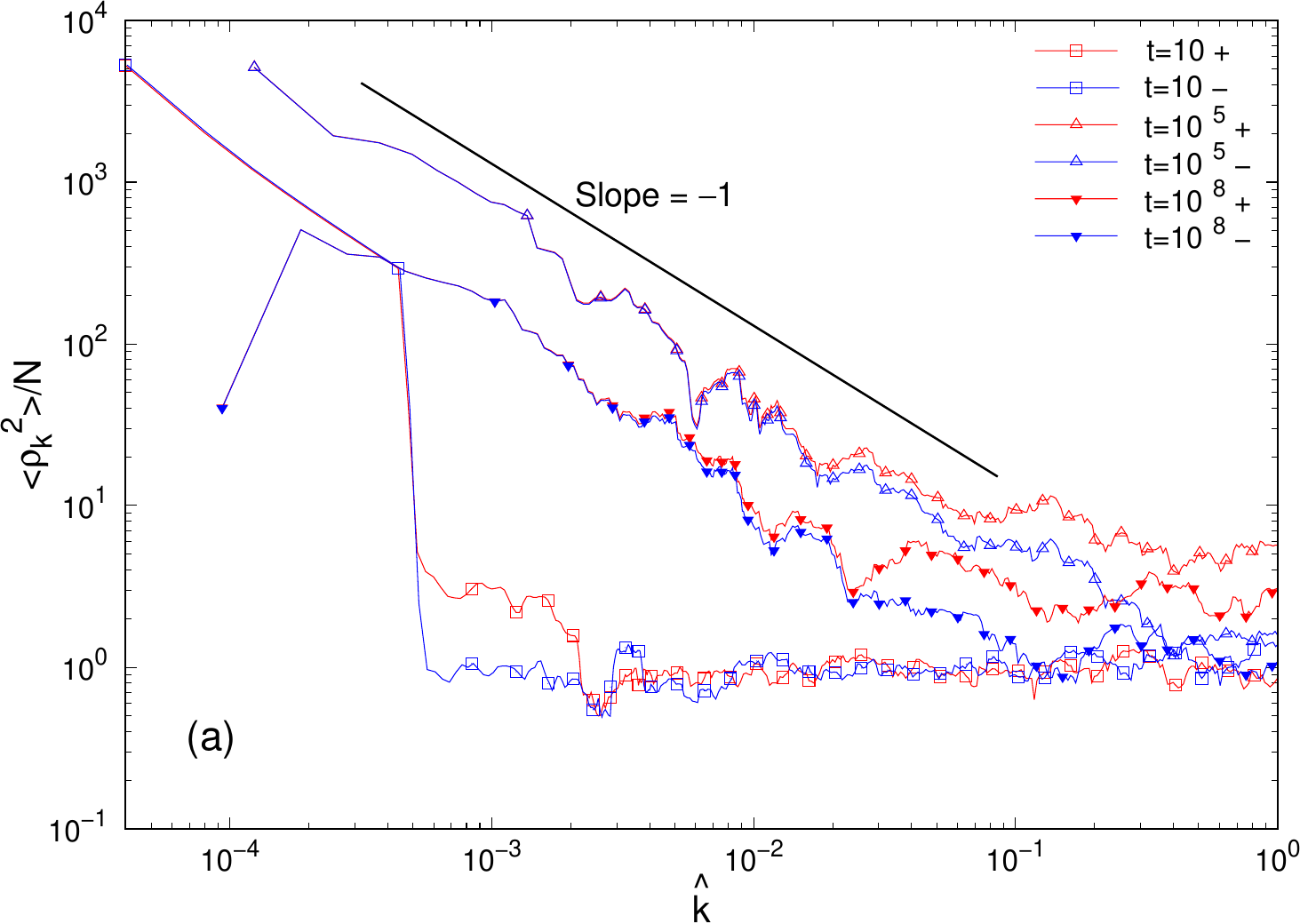} 
    \includegraphics[width=0.48\linewidth]{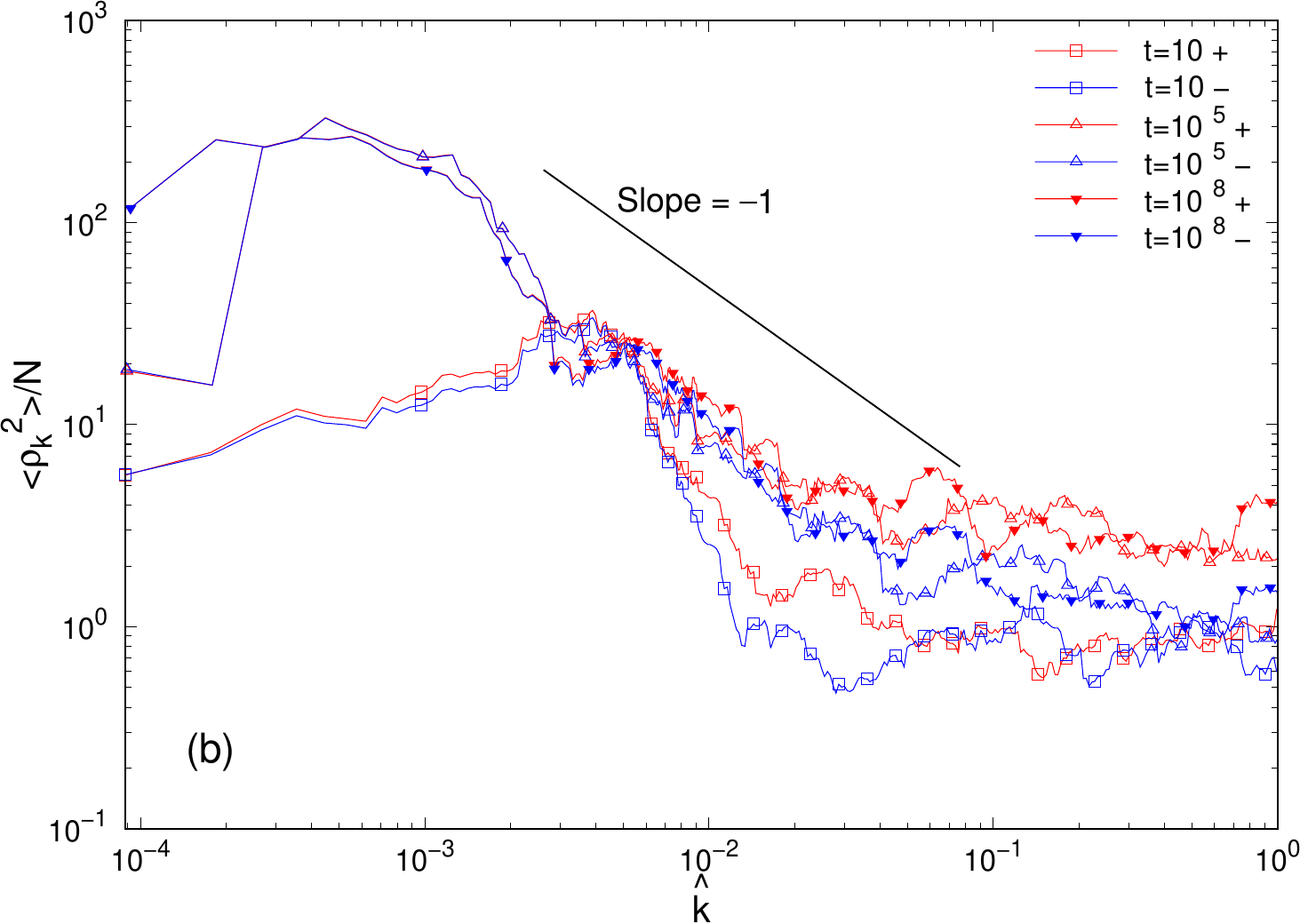} 
    \caption{Power spectra of the particle density $|\rho_{\hat k}|^2$ as a function of the scaled wave vector $\hat k$, at three different times, $t=10$ (squares), $t=10^5$ (upward triangles), and $t=10^8$ (downward triangles). Red lines refer to positive mass particles, and blue lines to negative mass particles. The initial velocity dispersion is $\sigma_{V}(0) = 10^4$ (a) and $\sigma_{V}(0) = 10$ (b). The straight black lines represents a spectrum proportional to $\hat k^{-1}$.}
    \label{fig:spectrum-scaled}
\end{figure}

For the same cases, Fig. \ref{fig:spectrum-scaled} shows the power spectra of the particle density $|\rho_k|^2$ as a function of the scaled wave vector in comoving co-ordinates. We note that the spectra of positive and negative mass particles are very similar, which is most probably a consequence of the formation of pairs (indeed, it can be interpreted as further evidence of the presence of such pairs). For the initially weakly coupled case (left panel), the late-time spectrum has a slope close to $-1$, which is somewhat steeper than the spectra observed in standard positive-mass Einstein-de Sitter simulations \cite{Miller2010,Miller2023}. This also appears to be true for the initially strongly coupled system, although the slope is less clear.
Interestingly, spectra of the type $|\rho_k|^2 \sim k^{-1}$ were observed in earlier simulations of an expanding universe, where the friction term in the comoving equations of motion had been removed (the so-called Hamiltonian model --- see Table 1 in Ref. \cite{Miller2010}).

\section{Discussion and conclusions}\label{sec:conclusion}

The Standard Cosmological Model ($\Lambda \rm CDM$) is a robust framework for describing the current state of our universe and its past history. One key component of the $\Lambda \rm CDM$ model is dark energy in the form of a cosmological constant $\Lambda$, which is responsible for driving the accelerated expansion of the universe. Such an acceleration, which was negligible in the early stages of the universe, becomes significant precisely at the present epoch, an intriguing fact known as the ``cosmological coincidence".
However, recent astronomical observations \cite{DES2025,DES-PRD} appear to indicate that the dark energy component is not actually constant, but diminishes over time. If confirmed, these findings, would have a profound impact on our understanding of the universe history.

In this work, we have identified and analyzed a new cosmological coincidence: the transition of the universe from a weakly coupled (collisionless) to a strongly coupled (collisional) gravitational regime. Remarkably, this coupling transition takes place at the same cosmic time as the standard cosmological coincidence. 

By considering a cosmological model containing equal amounts of positive and negative Bondi masses, we have shown that these two, apparently unrelated, coincidences can be understood as different manifestations of a single underlying mechanism. Although negative masses have been conjectured many times in the past, the scenario put forward by Bondi \cite{Bondi} -- where both the inertial and the gravitational (active and passive) masses are negative -- is the only one compatible with GR, with positive and negative particles following the same geodesics. In Newtonian terms, it is the only case that satisfies both the weak equivalence principle and Newton's law of action/reaction.
One of the peculiar features of Bondi masses is that pairs of positive/negative masses accelerate uniformly, while conserving their total energy and momentum (runaway acceleration). Such an effect may appear unphysical, but it has the potential of explaining the accelerated expansion of the universe without invoking dark energy or a cosmological constant. This is the main message of the present work.

In the weakly coupled regime, the dynamics of this mixed-mass system is well described by a Vlasov-Poisson mean-field model. Our linear response analysis demonstrates that such a system is always unstable, with growth rates that increase for shorter wavelengths. This feature, together with the cooling associated with the universe expansion, naturally drives the system toward a regime in which the assumption of collisionless dynamics breaks down, and two-body correlations become dominant. 

Once the system has become strongly coupled, the formation of stable positive/negative mass pairs triggers the Bondi runaway mechanism, which in turn produces a uniformly accelerating expansion. Long-time N-body simulations of the exact 1D Newtonian dynamics confirm this picture: after an initial coasting phase, followed by an intermediate regime characterized by random-walk acceleration, the system robustly settles into a phase where the root-mean-square displacement grows as $t^{2}$, signaling uniform acceleration. The onset of this accelerated phase coincides with the formation of strongly-coupled Bondi pairs, thereby linking the two cosmological transitions within a single physical framework. We have termed this particular positive/negative mass universe the ``Bondi universe".

These results suggest that cosmic acceleration may arise not from a dark-energy component or a cosmological constant, but from the intrinsic nonlinear dynamics of a mixed positive/negative mass universe transitioning into a strongly correlated state. Effectively, our model corresponds to a dark-energy component that is diluted by the universe expansion. However, and counter-intuitively, its impact does not wane with this dilution: quite to the contrary, its accelerating effect fully kicks in only when the mass density is low enough for two-body correlations to play a significant role via the Bondi runaway mechanism.
These features make our Bondi universe very different from other accelerating mechanisms, such as the backreaction scenario \cite{Schander2021,Green2014,Buchert2015}, which still assume a mean-field, weakly-coupled universe.

It is also important to stress the difference between the Bondi universe and other cosmological models that feature negative mass, such as the Dirac-Milne universe \cite{BenoitLevy,Chardin2018,Manfredi2018,Manfredi2020}. A universe containing equal amounts of positive and negative masses is gravitationally empty, and therefore behaves as a Milne universe \cite{Milne}, with the scale factor growing linearly in time, $a(t) \sim t$, which is usually referred to as a coasting universe \cite{Casado2020} \footnote{Note that, although empty universes are necessarily coasting, the reverse is not true, see \cite{Melia2012,John2019}.}.
However, the case of our Bondi universe is singularly different. Despite being gravitationally empty, it gives rise to a late acceleration stage where it transitions from a weakly-coupled to a strongly-coupled regime. Both the acceleration and the coupling transition occur at the same epoch, so that the Bondi model provides a common, unified explanation for both coincidences.

Finally, we point out that, while our model is idealized and relies on a 1D representation of Newtonian gravity, it captures essential physical mechanisms that should not be dimension-specific \cite{Miller2023}. 
Nonetheless, extending the present analysis to higher-dimensional settings and to relativistic frameworks remains a promising, and in some respects necessary, direction for future work. In particular, the precise details of Bondi pairs formation and the subsequent runaway acceleration may differ in 3D, because of the different behavior of Newton's force, which is spatially constant in 1D whereas it decays as the inverse squared distance in 3D. 
The role of angular momentum is also overlooked in a 1D geometry. Indeed, although the radial nature of Bondi pair interactions suggests that the runaway mechanism should still occur in 3D, the pair trajectories might exhibit greater intricacy and more sensitivity to perturbations.

In spite of the above limitations and the speculative hypothesis of negative mass, the results obtained in this work invite a re-examination of the standard interpretation of cosmic acceleration and open new avenues for exploring the large-scale dynamics of the universe without invoking dark energy.

\bibliography{bondi_biblio}

\appendix

\section{Virial theorem for an expanding self-gravitating system} \label{app:virial}

The virial theorem may be derived from the Vlasov-Poisson equations for standard (attractive) gravity
\begin{eqnarray}
\frac{\partial f}{\partial t} &+&
{\bm v}\cdot \nabla f - {\nabla \phi}  \cdot \nabla_{\bm v} f = 0, \label{app:vlasov} \\
\nabla^2 \phi &=& 4\pi G m n, \label{app:poisson}
\end{eqnarray}
where $f({\bm r},{\bm v},t)$ is the probability distribution in the phase space,  $n({\bm r},t)= \int f({\bm r},{\bm v},t) d{\bm v}$ is the number density, and $\phi({\bm r},t)$ is the gravitational potential. 

Following the standard derivation \cite{peebles2}, one obtains, in our notation:
\be
\frac{1}{2} \, \frac{d^2 }{dt^2} \,\langle |{\bm r}|^2 \rangle = 2K + U ,
\label{eq:virial}
\ee
where $\langle |{\bm r}|^2 \rangle \equiv \int \int f({\bm r},{\bm v},t) |{\bm r}|^2 d{\bm r} d{\bm v}$ is the moment of inertia per unit mass, and $K = {1 \over 2}\langle |{\bm v}|^2 \rangle$ and $U = {1 \over 2}\int n({\bm r},t) \phi({\bm r},t) d{\bm r}$ are, respectively, the kinetic and potential energies per unit mass.
Hence, for a stationary state: $2K + U =0$, which is the usual virial theorem.
Note that the quantities $\langle |{\bm r}|^2 \rangle$ and $\langle |{\bm v}|^2 \rangle$ correspond respectively to the variances labeled $\sigma_X^2$ and $\sigma_V^2$ in our 1D model, see Sec. \ref{sec:acceleration-A}.
We stress that they represent {\em global}, not local, variances; in particular, $\langle |{\bm v}|^2 \rangle$ is related to the total kinetic energy of the system, and does not represent a local peculiar velocity.

The virial theorem holds for confined self-gravitating systems at steady state, such as galaxies or clusters of galaxies, where the gravitational attraction dominates over the expansion of the universe. It does not, in principle, apply to the whole universe, which is not a stationary system because of the cosmological expansion. 
An extension of the virial theorem taking into account the expansion is provided by the Layzer-Irvine equation \cite{peebles2,Avelino2012}
\be
\frac{d}{dt}(K+U) + \frac{\dot a}{a}(2K+U) = 0.
\label{app:Layzer-Irvine}
\ee
However, the Layzer-Irvine equation is expressed in terms of comoving co-ordinates, whereas the present works uses a fixed co-ordinate system.

Therefore, we stick to the standard virial equation \eqref{eq:virial} and show how it should be interpreted for an expanding self-gravitating system, such as the one described by our numerical simulations. As discussed in Sec. \ref{sec:acceleration-A}, the root-mean-square displacement $\langle |{\bm r}|^2 \rangle^{1/2}$ is a good ersatz for the scale factor $a(t)$ as shown in Fig. \ref{fig:sigma204}.
Positing a power-law behavior for the scale factor: $a(t) =a_0 (t/t_0)^\beta$, one can write
\be
2K + U  = \beta (2\beta-1) \left(\frac{a_0}{t_0}\right)^2 \left(\frac{t}{t_0}\right)^{2\beta-2} .
\ee
We can deduce that:
\begin{itemize}
\item If $\beta >1$, then the virialization condition is never attained: $2K + U \to \infty$;
\item If $\beta =1$, then $2K + U = (a_0/t_0)^2 >0$, also no virialization;
\item If $\beta <1$, then $2K + U \to 0$, the system virializes at late times;
\item If $\beta = 1/2$, then $2K + U =0$ for all times.
\end{itemize}
Hence, virialization occurs only for expansions slower than coasting ($a \sim t$), such as the Einstein-de Sitter universe for which $a \sim t^{2/3}$. For the uniformly accelerated expansion considered in this work ($a \sim t^2$), or for an exponential expansion driven by a cosmological constant, the universe as a whole does not virialize. 
The case $\beta = 1/2$, corresponding to a radiation-dominated universe, is singular inasmuch as it remains strictly virialized for all times, provided it is so at $t=0$.

\section{Unstable evolution for a periodic system} \label{app:periodic}

In Sec. IV of the main text, based on a linear response analysis of the mean-field Vlasov-Poisson equations, it was shown that an admixture of positive and negative masses is always unstable, except in the special case where the two equilibrium distributions $f_{eq \pm}(v)$ are identical. The instability rate is proportional to the square root of the wave number, ${\rm Im} (\omega) \sim \sqrt{k}$, hence shorter wavelengths grow more rapidly.

\begin{figure}
    \centering
    \includegraphics[width=0.9\linewidth]{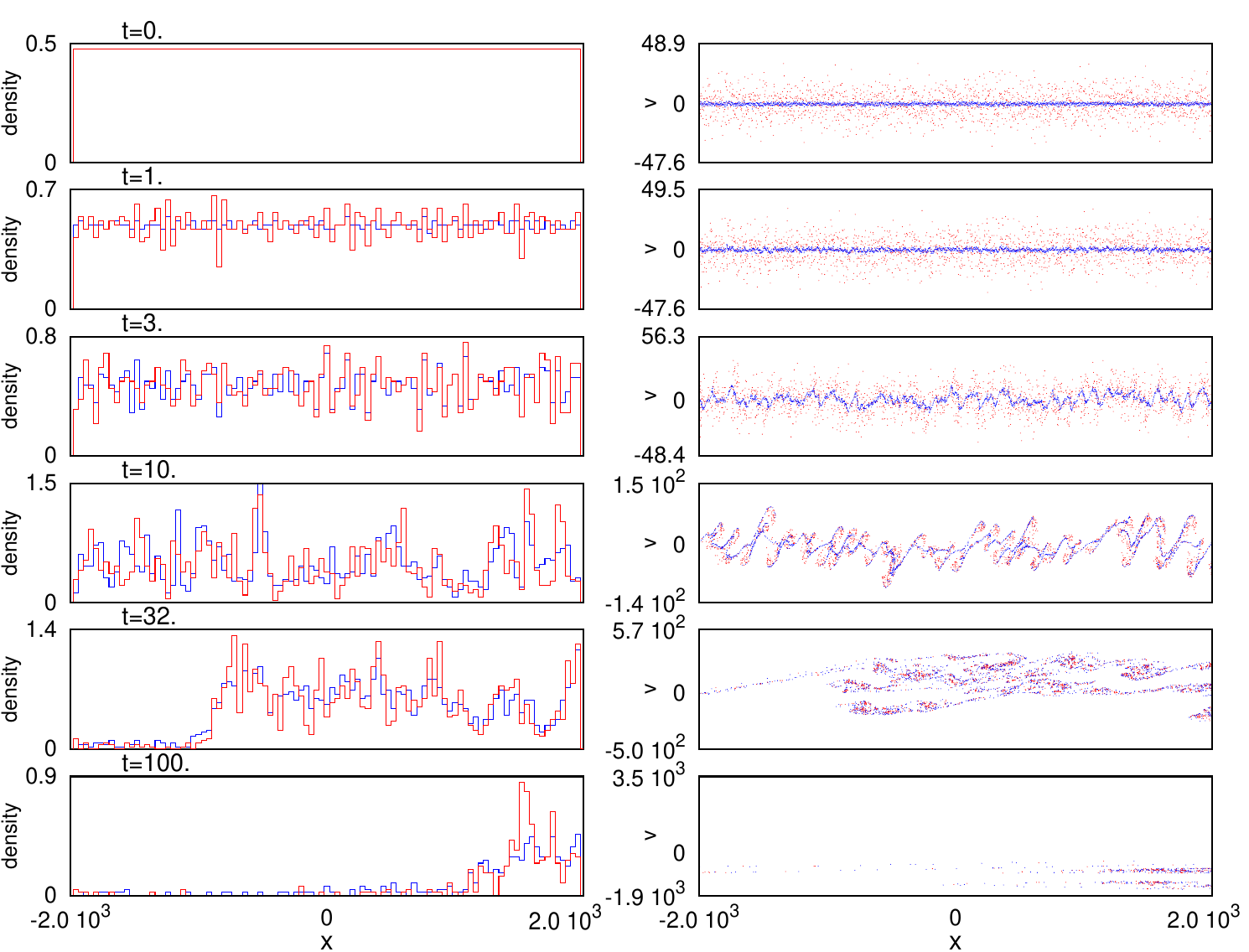}
    \caption{Periodic boundary conditions. Density plots (left panels) and phase space portraits (right panels) for positive mass particles (red) and negative mass particles (blue). The thermal velocities are  $v_{T-}=1$ and $v_{T+}=10$. The length of the domain is $L=N = 4 \times 10^3$.}
    \label{fig:suppl1}
\end{figure}

\begin{figure}
    \centering
    \includegraphics[width=0.9\linewidth]{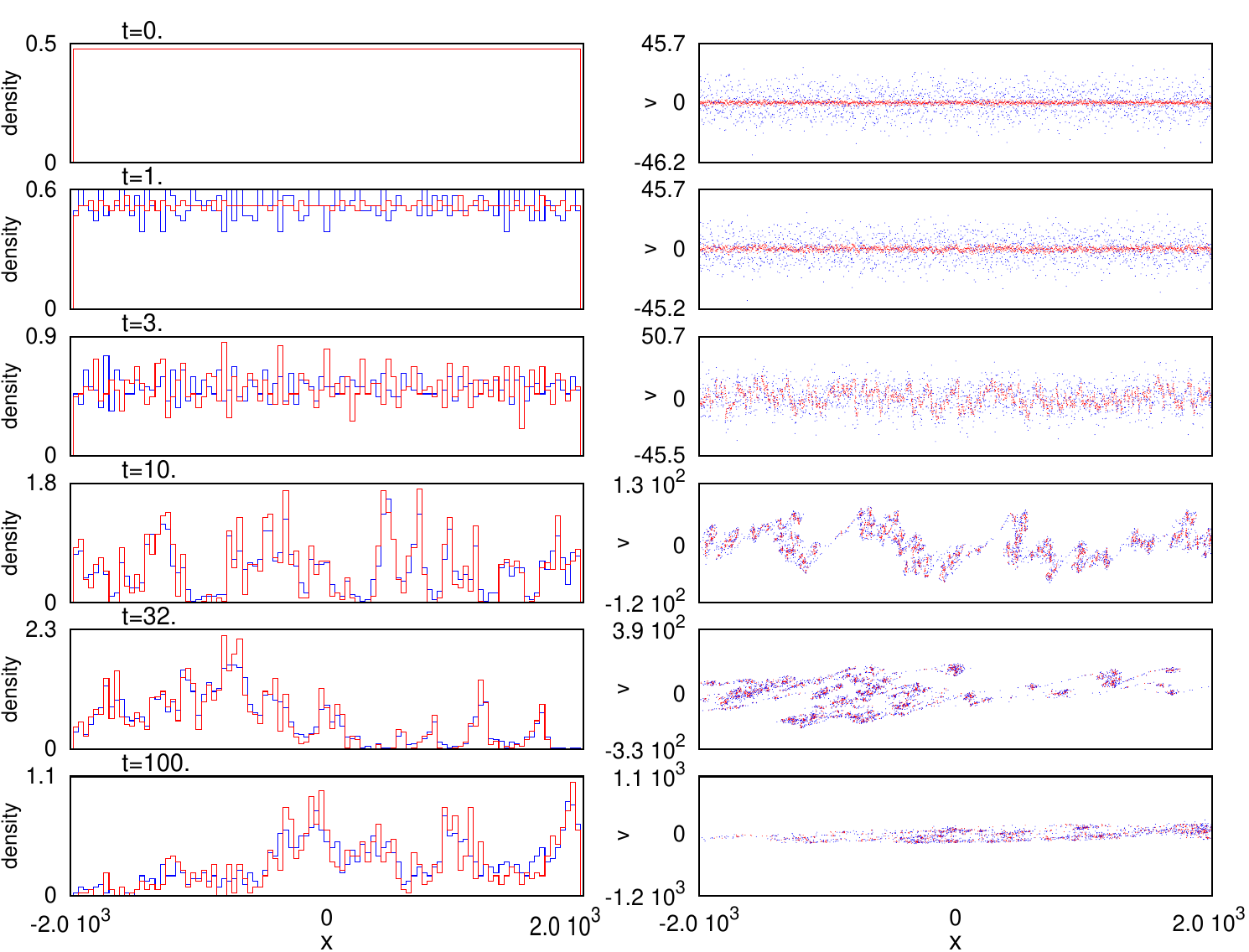}
    \caption{Periodic boundary conditions. Density plots (left panels) and phase space portraits (right panels) for positive mass particles (red) and negative mass particles (blue). The thermal velocities are  $v_{T-}=10$ and $v_{T+}=1$. The length of the domain is $L=N = 4 \times 10^3$.}
    \label{fig:suppl2}
\end{figure}

\begin{figure}
    \centering
    \includegraphics[width=0.9\linewidth]{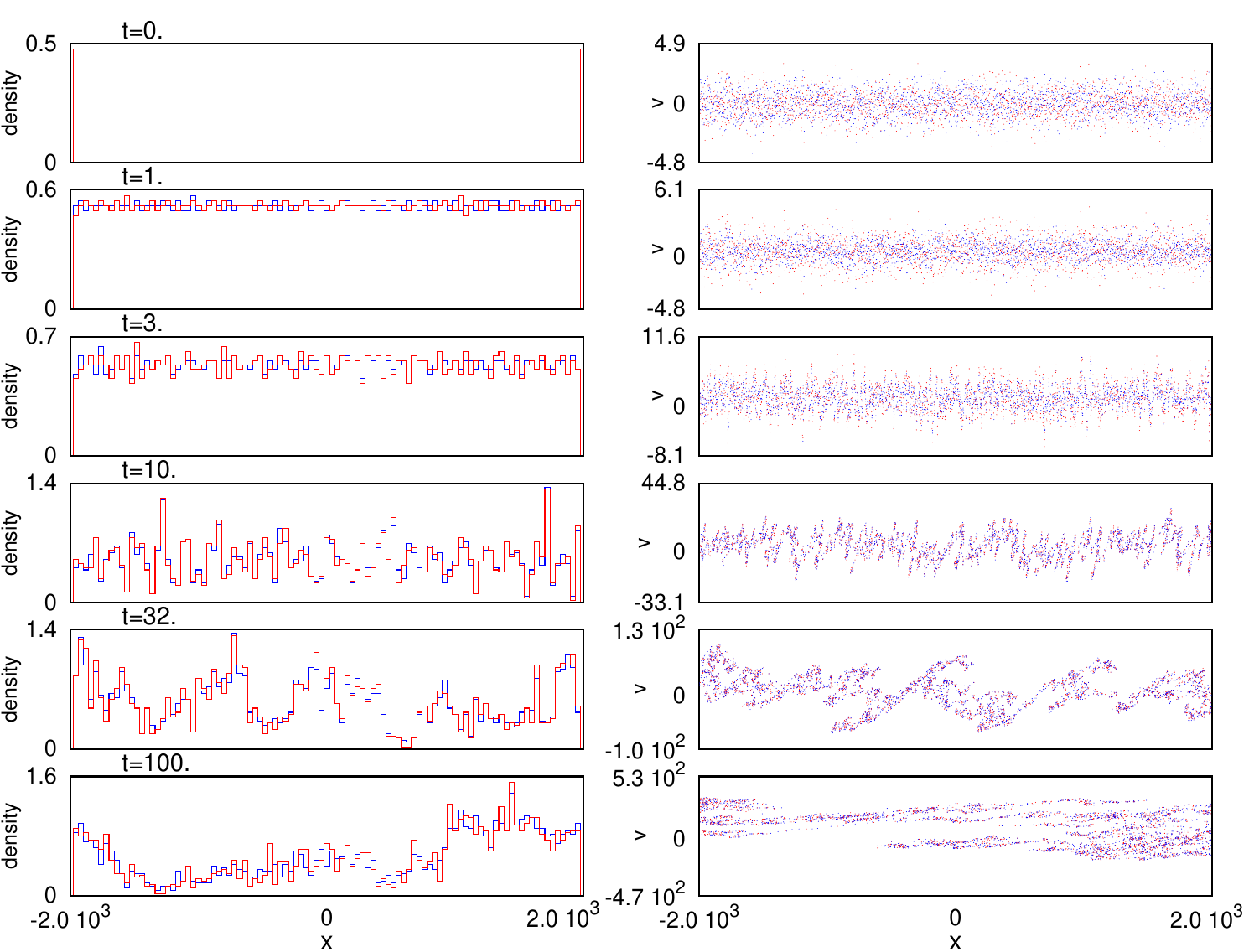}
    \caption{Periodic boundary conditions. Density plots (left panels) and phase space portraits (right panels) for positive mass particles (red) and negative mass particles (blue). The thermal velocities are  $v_{T-}=v_{T+}=1$. The length of the domain is $L=N = 4 \times 10^3$.}
    \label{fig:suppl3}
\end{figure}

Here, we show results of N-body numerical simulations  that confirm such instability. We use periodic boundary conditions with period $L=N = 4 \times 10^3$ and Maxwellian equilibria for both positive and negative mass species, with thermal velocities $v_{T \pm}$. We study three cases, for which the thermal velocities satisfy, respectively: $v_{T-} \ll v_{T+}$ (Fig. \ref{fig:suppl1}), $v_{T-} \gg v_{T+}$ (Fig. \ref{fig:suppl2}), and $v_{T-} = v_{T+}$ (Fig. \ref{fig:suppl3}).

For the two cases of unequal thermal velocities (Figs. \ref{fig:suppl1} and \ref{fig:suppl2}), the linear response theory predicts instability. As expected, short wavelengths quickly begin to grow during the linear phase ($t=0-3$). Later, nonlinearities lead to the formation of complex phase-space structures ($t=10-30$). For even longer times ($t \approx 100$), these structures become elongated and form filaments. 

The case $v_{T-} = v_{T+}$ (Fig. \ref{fig:suppl3}) should be stable according to the collisionless (mean field) linear response theory. However, the simulations were performed with the exact N-body code, which incorporates both collisionless and collisional effects. As the number of particles is finite, the two equilibrium Maxwellians are not exactly identical. Hence, the instability still occurs, although it is slower compared to the other cases. Indeed, in the equal thermal speed case of Fig. \ref{fig:suppl3}, fully nonlinear structures develop only around $t=32$, while they are visible as early as $t=10$ in the cases with different thermal speeds (Figs. \ref{fig:suppl1} and \ref{fig:suppl2}).


\section{Formation of Bondi pairs} \label{app:movie}

In a short movie (see below for the web link), we show the phase-space dynamics for a relatively small ($L=N =500$) and cold ($\sigma_V(0)=100$) system, which makes it easier to follow individual particles. Positive mass particles are depicted as red crosses and negative mass particles as blue squares.

The formation of a stable Bondi pair is evidenced by the fact that only the blue squares are visible, the red crosses being hidden behind them. The Bondi pairs become dominant around $t=2 \times 10^4$, when two streams of pairs are clearly visible. The two streams have opposite polarity (i.e., $+/-$ or $-/+$), and therefore accelerate either to the left or to the right.Occasionally, the particles in one pair switch positions, changing polarity, and start accelerating in the opposite direction, hence joining the other stream.
The simulation stops when no more particle crossings are observed. From that time on, the system evolves as a gas of independent Bondi pairs, each moving with constant acceleration in either direction.
\href{https://cloud.ipcms.fr/owncloud/s/SQJwV220TSBBKdh}{Watch the movie here.}

	\end{document}